\documentclass[aps]{revtex4}

\begin{document}
\title{Nonlinear Constraints from Non-Abelian Internal Symmetries}%
\author{Ludwik Turko}
\affiliation{Institute of Theoretical Physics, University of Wroclaw,\\
 pl. Maksa Borna 9, 50-204 Wroc\l aw, Poland}%
 \email{turko@ift.uni.wroc.pl}%
\date{Talk presented at August 29, 2001}
\begin{abstract}
Symmetry invariant local interaction of a many body system leads
to global constraints. We obtain explicit forms of the global
macroscopic condition assuring that at the microscopic level  the
evolution respects the overall symmetry.
\end{abstract}
\maketitle
\section{INTRODUCTION}

Let us consider a multiparticle interacting system with the internal
symmetry. The problem arises what are global constraints of the system
due to the symmetry conservation. In the microscopic formulation with
symmetry invariant dynamical equations the answer is given by an
analysis of corresponding solutions --- assuming that solutions are
known.

We are looking here for global conditions to provide consistency with
the overall symmetry of the system. These conditions should not depend
an exact analytic form of the solutions. An old fashioned example would
be Kepler's laws in the classical mechanics which are related to the
orbital momentum conservation and can be proved without knowledge of the
analytic solutions of Newton equations. As an another very simple
example let  us consider $N^{(a)}, N^{(b)},\dots, N^{(n)}$ charged
particles with corresponding individual charges $q_a, q_b,\dots, q_n.$
Particle numbers are time-dependent but the global charges must be
conserved (exact $U(1)$ symmetry). So there is a condition
\begin{equation}\label{abcharge}
q_a\frac{d N^{(a)}}{dt} + q_b\frac{d
N^{(b)}}{dt}+ \cdots + q_n\frac{d N^{(n)}}{dt} = 0\,,
\end{equation}
valid for any charge conserving interaction.

Our aim\cite{TurRaf} is to find a corresponding condition for non-abelian
symmetries.

\section{GENERALIZED PROJECTION METHOD}

On the first Bielefeld conference in 1980 I presented a paper
\cite{Redlich:1980bf} devoted to the problem of a formulation of
thermodynamics with internal symmetries taken into account. Group
projection techniques allowed for a consistent treatment of
equilibrium systems and gave tools to obtain canonical partition
functions corresponding to the system transforming under given
representation of the symmetry group. This technique can also be
used for a more general non-static problem.

Let us consider a system consisting of particles belonging to
multiplets $\alpha_j$ of the symmetry group. Particles from the
given multiplet $\alpha_j$ are characterized by quantum numbers
$\nu_j$ --- related to the symmetry group, and quantum numbers
$\zeta_j$ characterizing different multiplets of the same irreducible
representation $\alpha_j$.

The number of particles of the specie
$\{\alpha,\nu_\alpha;\zeta\}$ is denoted here by
$N^{(\alpha)}_{\nu_\alpha;(\zeta)}.$ These occupation numbers are
time dependent until the system reaches the chemical equilibrium.
However the representation of the symmetry group for the system
remains constant in the course of a time evolution. A multiplicity
$N^{(\alpha_j)}$ of the representation $\alpha_j$ in this product
is equal to a number of particles which transform under this
representation:
\begin{equation}
N^{(\alpha_j)} = \sum_j\left(\sum_{\zeta_j}\,
N^{(\alpha_j)}_{\nu_{\alpha_j};(\zeta_j)}\right)= \sum_j\,
N^{(\alpha_j)}_{\nu_{\alpha_j}}\,.
\end{equation}
We introduce a state vector $\left\vert
N^{(\alpha_1)}_{\nu_{\alpha_1}},\dots,
N^{(\alpha_n)}_{\nu_{\alpha_n}}\right\rangle $ in particle number
representation. The probability that
$N^{(\alpha_1)}_{\nu_{\alpha_1}},\dots, N^{(\alpha_n)}_{
\nu_{\alpha_n}}$ particles transforming under the symmetry group
representations $\alpha_1,\dots,\alpha_n$ combine into a state
transforming under representation $\Lambda$ of the symmetry group
is given by
\begin{eqnarray}\label{proj1states}
\overline{P^{\Lambda,\lambda_{\Lambda}}_{\{N^{(\alpha_1)}_{\nu_{\alpha_1}},
\dots,\,N^{(\alpha_n)}_{\nu_{\alpha_n}}\}}}\quad =
\quad\left\langle N^{(\alpha_1)}_{\nu_{\alpha_1}},
\cdots,N^{(\alpha_n)}_{\nu_{\alpha_n}}\right\vert
{\mathcal P}^{\Lambda}\left\vert N^{(\alpha_1)}_{\nu_{\alpha_1}},
\dots, N^{(\alpha_n)}_{\nu_{\alpha_n}}\right\rangle
\end{eqnarray}
The projection operator ${\mathcal P}^{\Lambda}$ has the form
(see e.g. \cite{Wigner}):
\begin{equation}\label{proj1}
{\mathcal
P}^{\Lambda}=d(\Lambda)\int\limits_G\,d\mu(g)\bar\chi^{(\Lambda)}(g)U(g)\,.
\end{equation}
Here $\chi^{(\Lambda)}$ is the character of the representation
$\Lambda$, $d(\Lambda)$ is the dimension of the representation,
$d\mu(g)$ is the invariant Haar measure on the group, and $U(g)$
is an operator transforming a state under consideration. In
particle number representation the operator $U(g)$ is defined as:
\begin{eqnarray}\label{transf}
\lefteqn{U(g)\left\vert N^{(\alpha_1)}_{\nu_{\alpha_1}},\dots,
N^{(\alpha_n)}_{\nu_{\alpha_n}}\right\rangle}\\ & &
=\sum\limits_{\nu_1^{(1)},\dots,\nu_n^{(N_{\nu_n})}}\,
D^{(\alpha_1)}_{\nu_1^{(1)}\nu_1}\!\!\cdots
D^{(\alpha_1)}_{\nu_1^{(N_{\nu_1})}\nu_1}\!\!\cdots
D^{(\alpha_n)}_{\nu_n^{(1)}\nu_n}\cdots
D^{(\alpha_n)}_{\nu_n^{(N_{\nu_n})}\nu_n} \left\vert
N^{(\alpha_1)}_{\nu_{\alpha_1}},\dots,
N^{(\alpha_n)}_{\nu_{\alpha_n}}\right\rangle\,.\nonumber
\end{eqnarray}
$D^{(\alpha_n)}_{\nu,\nu}$ is a matrix elements of the group
element $g$ corresponding to the representation $\alpha$.

One gets finally
\begin{equation} \label{weights}
\overline{P^{\Lambda,\lambda_{\Lambda}}_{\{N^{(\alpha_1)}_{\nu_{\alpha_1}},
\dots,\,N^{(\alpha_n)}_{\nu_{\alpha_n}}\}}}\quad = \quad {\mathcal
A}^{\{{\mathcal N}\}}
d(\Lambda)\int\limits_G\,d\mu(g)\bar\chi^{(\Lambda)}(g)
[D^{(\alpha_1)}_{\nu_1\nu_1}]^{N^{(\alpha_1)}_{\nu_{\alpha_1}}}
\cdots
[D^{(\alpha_n)}_{\nu_n\nu_n}]^{N^{(\alpha_n)}_{\nu_{\alpha_n}}}\,.
\end{equation}
$D^{(\alpha_n)}_{\nu,\nu}$ is a matrix elements of the group
element $g$ corresponding to the representation $\alpha$ and
${\mathcal A}^{\{{\mathcal N}\}}$ is an overall permutation
normalization factor:
\begin{equation}\label{faktor}
{\mathcal A}^{\{{\mathcal
N}\}}= \prod\limits_j\prod_{\zeta_j}{\mathcal
A}^{\alpha_j}_{(\zeta_j)}\,,
\end{equation}
where ${\mathcal A}^{\alpha_j}_{(\zeta_j)}$ are partial factors for
particles of the kind $\{\alpha,\zeta\}:$
\begin{eqnarray}\label{faktorp}
{\mathcal A}^\alpha_{(\zeta)}= \frac{{\mathcal
N}^{(\alpha)}_{(\zeta)}!}{\prod\limits_{\nu_\alpha} {\mathcal
N}^{(\alpha)}_{\nu_\alpha;(\zeta)}!}\ ; \qquad
\end{eqnarray}
The permutation factor gives a proper normalization of state vectors
reflecting indistinguishability of particles:
\begin{equation}\label{norma}
\left\langle N^{(\alpha_1)}_{\nu_{\alpha_1}},
\cdots,N^{(\alpha_n)}_{\nu_{\alpha_n}}\right\vert \left.
N^{(\alpha_1)}_{\nu_{\alpha_1}},\dots,
N^{(\alpha_n)}_{\nu_{\alpha_n}}\right\rangle={\cal A}^{\{{\cal
N}\}}\ ;
\end{equation}
Because of symmetry conservations all weights in
Eq.\,(\ref{weights}) should be constant:
\begin{equation} \label{cond}
\frac{d}{dt}\overline{P^{\Lambda,\lambda_{\Lambda}}_{\{N^{(\alpha_1)}_{\nu_{\alpha_1}},
\dots,\,N^{(\alpha_n)}_{\nu_{\alpha_n}}\}}}\quad =\quad 0 .
\end{equation}
Introducing  here the result of Eq.\,(\ref{weights}) one obtains:
\begin{eqnarray} \label{deriv}
\lefteqn{0 = \frac{d\,{\mathcal A}^{\{{\mathcal
N}\}}}{dt}d(\Lambda)\int\limits_G d\mu(g)\bar\chi^{(\Lambda)}(g)
[D^{(\alpha_1)}_{\nu_1\nu_1}]^{N^{(\alpha_1)}_{\nu_{\alpha_1}}}\cdots
[D^{(\alpha_n)}_{\nu_n\nu_n}]^{N^{(\alpha_n)}_{\nu_{\alpha_n}}}}\\ & &
+\sum_{j=1}^n\sum_{\nu_{\alpha_j}}\,
\frac{d\,N^{(\alpha_j)}_{\nu_{\alpha_j}}}{dt} {\mathcal
A}^{\{{\mathcal N}\}}d(\Lambda)\int\limits_G
d\mu(g)\bar\chi^{(\Lambda)}(g)
[D^{(\alpha_1)}_{\nu_1\nu_1}]^{N^{(\alpha_1)}_{\nu_{\alpha_1}}}
\cdots
[D^{(\alpha_n)}_{\nu_n\nu_n}]^{N^{(\alpha_n)}_{\nu_{\alpha_n}}}
\log[D^{(\alpha_j)}_{\nu_j\nu_j}] .\nonumber
\end{eqnarray}

The integrals which appear in Eq.\,(\ref{deriv}) can be expressed
explicitly in an analytic form for any compact symmetry group.

To write an expression for the time derivative of the
normalization factor ${\mathcal A}^{\{{\mathcal N}\}}$ we perform
analytic continuation from integer to continuous values of
variables $N^{(\alpha_n)}_{\nu_{\alpha_n}}.$ All factorials in
Eq.\, (\ref{faktor}) are replaced by the $\Gamma$--function  of
corresponding arguments.  We encounter here also the digamma
function $\psi$ \cite{Abram}:
\begin{equation}\label{digamma}
\psi(x)=\frac{d\, \log\Gamma(x)}{d\,x} .
\end{equation}
This allows to write:
\begin{equation}
\frac{d\,{\mathcal A}^{\{{\mathcal N}\}}}{dt}
 = {\mathcal A}^{\{{\mathcal N}\}}\sum_j\sum_{\zeta_j}
 \left[\frac{d\,{\mathcal N}^{(\alpha_j)}_{(\zeta_j)}}{dt}
 \psi({\mathcal N}^{(\alpha_j)}_{(\zeta_j)}+1)
 -\sum_{\nu_{\alpha_j}}
\frac{d\,{\mathcal N}^{(\alpha_j)}_{\nu_{\alpha_j};(\zeta_j)}}{dt}
\psi({\mathcal N}^{(\alpha_j)}_{\nu_\alpha;(\zeta_j)}+1)\right]
 \label{evfactor}
\end{equation}
Eq.\,(\ref{deriv}) can be written in a form
\begin{eqnarray}\label{derisimpl}
\lefteqn{\sum_{j=1}^n\sum_{\nu_{\alpha_j}}\,
\frac{d\,N^{(\alpha_j)}_{\nu_{\alpha_j}}}{dt} \frac{d\,\log
\widetilde
\mathcal{P}^{\Lambda,\lambda_{\Lambda}}_{\{N^{(\alpha_1)}_{\nu_{\alpha_1}},
\dots,\,N^{(\alpha_n)}_{\nu_{\alpha_n}}\}}}
{d\,N^{(\alpha_j)}_{\nu_{\alpha_j}}}}\\ & &
 = \sum_j\sum_{\zeta_j}
\left(-\frac{d\,N^{(\alpha_j)}_{(\zeta_j)}}{dt}
\psi(N^{(\alpha_j)}_{(\zeta_j)}+1)+\sum_{\nu_{\alpha_j}}
\frac{d\,N^{(\alpha_j)}_{\nu_{\alpha_j};(\zeta_j)}}{dt}
\psi(N^{(\alpha_j)}_{\nu_{\alpha_j};(\zeta_j)}+1)\right)\nonumber
\, .
\end{eqnarray}
where
\begin{equation}\label{otherP}
\widetilde {\mathcal
P}^{\Lambda,\lambda_{\Lambda}}_{\{N^{(\alpha_1)}_{\nu_{\alpha_1}},
\dots,\,N^{(\alpha_n)}_{\nu_{\alpha_n}}\}}=\int\limits_G
d\mu(g)\bar\chi^{(\Lambda)}(g)
[D^{(\alpha_1)}_{\nu_1\nu_1}]^{N^{(\alpha_1)}_{\nu_{\alpha_1}}}
\cdots
[D^{(\alpha_n)}_{\nu_n\nu_n}]^{N^{(\alpha_n)}_{\nu_{\alpha_n}}}
\end{equation}
is analytically extended for continuous values of variables
$N^{(\alpha_j)}_{\nu_{\alpha_j}}.$ This gives
\begin{equation}\label{derivP}
\frac{d\,\log \widetilde
\mathcal{P}^{\Lambda,\lambda_{\Lambda}}_{\{N^{(\alpha_1)}_{\nu_{\alpha_1}},
\dots,\,N^{(\alpha_n)}_{\nu_{\alpha_n}}\}}}
{d\,N^{(\alpha_j)}_{\nu_{\alpha_j}}}\quad =\quad
\frac{\int\limits_G d\mu(g)\bar\chi^{(\Lambda)}(g)
[D^{(\alpha_1)}_{\nu_1\nu_1}]^{N^{(\alpha_1)}_{\nu_{\alpha_1}}}
\cdots
[D^{(\alpha_n)}_{\nu_n\nu_n}]^{N^{(\alpha_n)}_{\nu_{\alpha_n}}}
\log[D^{(\alpha_j)}_{\nu_j\nu_j}]}{\int\limits_G
d\mu(g)\bar\chi^{(\Lambda)}(g)
[D^{(\alpha_1)}_{\nu_1\nu_1}]^{N^{(\alpha_1)}_{\nu_{\alpha_1}}}
\cdots
[D^{(\alpha_n)}_{\nu_n\nu_n}]^{N^{(\alpha_n)}_{\nu_{\alpha_n}}}}\,.
\end{equation}
For higher multiplicities digamma functions in
Eq.\,(\ref{derisimpl}) can be replaced by corresponding
logarithmic functions according to the asymptotic
formula\cite{Abram}:
\begin{equation}\label{asymptdigamma}
\psi(N+1)\approx\log N\,.
\end{equation}
Eqs.\,(\ref{derisimpl}) and (\ref{derivP}) give a set of
conditions related to the internal symmetry of a system. They are
meaningful only for nonzero values of coefficients
(\ref{weights}). It is easy to see that coefficients
$\overline{P^{\Lambda,\lambda_{\Lambda}}_{\{N^{(\alpha_1)}_{\nu_{\alpha_1}},
\dots,\,N^{(\alpha_n)}_{\nu_{\alpha_n}}\}}}$ are different from
zero only if parameters $\lambda_{\Lambda}$ are consistent with
the conservation of the simultaneously measurable charges related
to the symmetry group. A number of such charges is equal to the
rank $k$ of the symmetry group. For the isospin $SU(2)$ group that
is the third component of the isospin, for the flavour $SU(3)$
that would be the third component of the isospin and the
hypercharge. In general case one has $k$ linear relations between
variables $N^{(\alpha_j)}_{\nu_{\alpha_j}}$ what reduces
correspondingly the number of independent variables.

\section{CONCLUSIONS}

We have got relations which are {\it necessary} global conditions
to provide consistency with the overall symmetry of the system.
They do not depend on the form of the underlying microscopic
interaction.  Abelian internal symmetries lead to simple and
obvious linear relations as in Eq.\,(\ref{abcharge}). Non-abelian
internal symmetries lead to nonlinear relations as in
Eq.\,(\ref{deriv}).

If we know solutions of symmetry invariant evolution equations then all
those constraints would become identities. In other case they give a
subsidiary information about the system and can be  used as a
consistency check for approximative calculations. A case of generalized
Vlasov - Boltzman kinetic equations was considered in \cite{TurRaf}.

New constraints lead to decreasing number of available states for
the system during its evolution. New correlations appear and the
change in the thermodynamical behavior can be expected
\cite{Elze:2001ss}.

A challenging point is to find structures which would correspond
to chemical potentials when system approaches the equilibrium
distribution. The equilibrium distribution in the presence of
constraints can be constructed by the Lagrange multipliers method.
The multipliers related to the ``abelian constraints, such as
Eq.\,(\ref{abcharge}), are well known chemical potentials.
Multipliers related to the ``nonabelian" constraints (\ref{deriv})
are more complicated.  Because these constraints are nonlinear
ones, corresponding multipliers cannot be treated as standard
additive thermodynamical potentials.

\begin{acknowledgments}
Work supported in part by the Polish Committee for Scientific
Research under contract KBN~-~2P03B~030~18
\end{acknowledgments}

\end{document}